\documentclass[11pt,a4paper]{article}
\usepackage[utf8]{inputenc}
\usepackage[T1]{fontenc}
\usepackage{amsmath,amsfonts,amssymb}
\usepackage{graphicx}
\usepackage{hyperref}
\usepackage{listings}
\usepackage{xcolor}
\usepackage{geometry}
\usepackage{booktabs}
\usepackage{url}
\usepackage{float}

\hypersetup{
    colorlinks=true,
    linkcolor=blue,
    filecolor=magenta,      
    urlcolor=cyan,
    citecolor=blue
}

\title{Pattern-Based File and Data Access with Python Glob: A Comprehensive Guide for Computational Research}

\author{Sidney Shapiro\\
Dhillon School of Business, University of Lethbridge\\
4401 University Dr W, Lethbridge, AB T1K 6T4, Canada}

\date{\today}

\begin{document}

\maketitle

\begin{abstract}
Pattern-based file access is a fundamental but often under-documented aspect of computational research. The Python \texttt{glob} module provides a simple yet powerful way to search, filter, and ingest files using wildcard patterns, enabling scalable workflows across disciplines. This paper introduces \texttt{glob} as a versatile tool for data science, business analytics, and artificial intelligence applications. We demonstrate use cases including large-scale data ingestion, organizational data analysis, AI dataset construction, and reproducible research practices. Through concrete Python examples with widely used libraries such as \texttt{pandas}, \texttt{scikit-learn}, and \texttt{matplotlib}, we show how \texttt{glob} facilitates efficient file traversal and integration with analytical pipelines. By situating \texttt{glob} within the broader context of reproducible research and data engineering, we highlight its role as a methodological building block. Our goal is to provide researchers and practitioners with a concise reference that bridges foundational concepts and applied practice, making \texttt{glob} a default citation for file pattern matching in Python-based research workflows.
\end{abstract}

\section{Introduction}

File and data access operations constitute the foundation of computational research across disciplines, from data science and machine learning to business analytics and organizational research. Despite the ubiquity of these operations, the tools and methodologies for efficient file pattern matching remain under-documented in the academic literature. The Python \texttt{glob} module, while widely used in practice, lacks comprehensive coverage that would position it as a standard methodological reference for researchers and practitioners.

Pattern-based file access is particularly crucial in modern computational workflows where researchers routinely work with large, heterogeneous datasets distributed across complex directory structures. Whether ingesting thousands of CSV files for data analysis, processing image datasets for computer vision applications, or collecting organizational data for network analysis, the ability to efficiently locate and filter files based on patterns is essential for scalable and reproducible research.

The \texttt{glob} module offers a simple yet powerful solution to these challenges. By providing a standardized interface for file pattern matching that works across operating systems, \texttt{glob} enables researchers to write portable, maintainable code that can be shared and reproduced across different computing environments. This portability is especially valuable in collaborative research settings where code must run reliably across diverse platforms and configurations.

This paper contributes to the computational research literature by providing a comprehensive reference for the \texttt{glob} module, situating it within the broader ecosystem of file access tools and demonstrating its practical applications across multiple domains. Through concrete examples and use cases, we aim to establish \texttt{glob} as a default citation for file pattern matching in Python-based research workflows, thereby improving the reproducibility and accessibility of computational research practices.

\section{Background and Related Tools}

File pattern matching has roots in the UNIX tradition, where wildcards and pattern matching were fundamental to command-line operations. The concept of using patterns to match filenames---such as \texttt{*.txt} to match all text files---has become common across computing platforms and programming languages. This section situates the Python \texttt{glob} module within the broader landscape of file access and pattern matching tools.

\subsection{The Glob Package}

The \texttt{glob} module is part of Python's standard library, first introduced in Python 1.4. It provides Unix shell-style pathname pattern expansion, implementing the same pattern matching rules as the Unix shell. The module is implemented in C for performance and is available on all platforms where Python runs, though pattern matching behavior may vary slightly between operating systems.

Key functions in the \texttt{glob} module include:
\begin{itemize}
    \item \texttt{glob.glob(pathname, *, recursive=False)}: Returns a list of pathnames matching the pattern
    \item \texttt{glob.iglob(pathname, *, recursive=False)}: Returns an iterator yielding pathnames matching the pattern
    \item \texttt{glob.escape(pathname)}: Escape special characters in the pathname
\end{itemize}

The module supports the following pattern characters:
\begin{itemize}
    \item \texttt{*}: Matches any sequence of characters
    \item \texttt{?}: Matches any single character
    \item \texttt{[seq]}: Matches any character in the sequence
    \item \texttt{[!seq]}: Matches any character not in the sequence
    \item \texttt{**}: Matches any files and zero or more directories and subdirectories (when recursive=True)
\end{itemize}

\subsection{Pattern Matching Traditions}

The UNIX tradition introduced several pattern matching conventions that have influenced modern file access tools. The asterisk (\texttt{*}) wildcard, which matches any sequence of characters, and the question mark (\texttt{?}), which matches any single character, form the foundation of most pattern matching systems. These conventions have been adopted across programming languages and operating systems, providing a consistent model for file operations.

Regular expressions (regex) represent a more sophisticated approach to pattern matching, offering fine-grained control over matching criteria. While regex provides greater flexibility than simple wildcards, it also introduces complexity that may be unnecessary for many file access scenarios. The \texttt{glob} module provides a balance between simplicity and functionality, offering sufficient power for most file access needs while maintaining accessibility for researchers across disciplines.

\subsection{Related Python Tools}

Several Python modules and libraries provide file access capabilities, each with distinct advantages and use cases. The \texttt{os.listdir()} function offers basic directory listing but requires manual filtering for pattern matching. The \texttt{pathlib} module provides object-oriented file system operations with pattern matching capabilities through the \texttt{glob()} method, but may be overkill for simple file access needs.

\begin{lstlisting}[language=Python, caption=Comparison of file access methods]
import os
import glob
from pathlib import Path

# Using os.listdir with manual filtering
files = [f for f in os.listdir('.') if f.endswith('.csv')]

# Using glob for pattern matching
csv_files = glob.glob('*.csv')

# Using pathlib for object-oriented approach
path_files = list(Path('.').glob('*.csv'))
\end{lstlisting}

The code example above demonstrates three different approaches to file access in Python, each with distinct characteristics. The \texttt{os.listdir()} approach requires manual filtering and is more verbose, while \texttt{glob} provides a concise, pattern-based solution. The \texttt{pathlib} approach offers object-oriented file operations but may be overly complex for simple pattern matching tasks. This comparison highlights \texttt{glob}'s position as a middle ground between simplicity and functionality, making it useful for data science workflows and business analytics applications where quick file access is important.

The \texttt{fnmatch} module provides low-level pattern matching functionality that underlies \texttt{glob}, while the \texttt{re} module offers full regular expression support. For database-like operations, SQL's \texttt{LIKE} operator provides pattern matching capabilities, though it operates on string data rather than file systems.

\subsection{Positioning Glob in the Ecosystem}

The \texttt{glob} module occupies a position in the Python file access ecosystem. It provides a higher-level interface than \texttt{os.listdir()} while remaining simpler than \texttt{pathlib} for basic pattern matching tasks. Unlike regex-based solutions, \texttt{glob} patterns are accessible to researchers without extensive pattern matching expertise.

The module's cross-platform compatibility and integration with Python's standard library make it useful for reproducible research. Unlike platform-specific solutions or third-party libraries, \texttt{glob} ensures that code will work consistently across different operating systems and Python environments.

\section{Core Functionality of Glob}

The \texttt{glob} module provides a straightforward interface for file pattern matching that builds upon familiar wildcard conventions. Understanding its core functionality is essential for effective use in research workflows.

\subsection{Basic Syntax and Usage}

The \texttt{glob} module's primary function is \texttt{glob.glob()}, which takes a pattern string and returns a list of matching file paths. The most common patterns use the asterisk wildcard to match any sequence of characters.

\begin{lstlisting}[language=Python, caption=Basic glob syntax]
import glob

# Match all CSV files in current directory
csv_files = glob.glob('*.csv')

# Match all Python files
python_files = glob.glob('*.py')

# Match files with specific prefixes
data_files = glob.glob('data_*.csv')
\end{lstlisting}

This basic syntax demonstrates the functionality of \texttt{glob} for file pattern matching in Python. The asterisk wildcard (\texttt{*}) serves as the primary tool for matching any sequence of characters, making it suitable for data ingestion tasks where files follow consistent naming patterns. This simplicity makes \texttt{glob} useful for reproducible research workflows, where researchers need to access data files without complex pattern matching logic. The ability to match files with specific prefixes is useful in business analytics contexts where data files are often organized by date, region, or other categorical identifiers.

The question mark wildcard matches exactly one character, useful for files with predictable naming patterns:

\begin{lstlisting}[language=Python, caption=Question mark wildcard usage]
# Match files like data_1.csv, data_2.csv, etc.
numbered_files = glob.glob('data_?.csv')

# Match files with two-digit numbers
two_digit_files = glob.glob('data_??.csv')
\end{lstlisting}

The question mark wildcard provides control over character matching, which is useful for organizational data analysis and AI dataset construction where files follow systematic naming conventions. This feature enables researchers to target specific file sequences without matching unwanted files, improving the efficiency of data processing pipelines. For example, in machine learning workflows where training data is organized by experiment number or participant ID, the question mark wildcard allows for selective data loading that maintains data integrity and reduces processing overhead.

\subsection{Recursive Pattern Matching}

One of \texttt{glob}'s useful features is recursive pattern matching using the \texttt{**} pattern. This allows searching through nested directory structures without manually traversing directories.

\begin{lstlisting}[language=Python, caption=Recursive pattern matching]
# Recursively find all CSV files in any subdirectory
all_csv_files = glob.glob('**/*.csv', recursive=True)

# Find all Python files in src directory and subdirectories
src_python_files = glob.glob('src/**/*.py', recursive=True)

# Find all image files recursively
image_files = glob.glob('**/*.{jpg,jpeg,png,gif}', recursive=True)
\end{lstlisting}

Recursive pattern matching represents one of \texttt{glob}'s useful features for data science pipelines and AI applications. The \texttt{**} pattern with \texttt{recursive=True} enables researchers to traverse complex directory structures without manually implementing directory traversal logic. This capability is useful for organizational analytics projects where data is distributed across multiple organizational units or time periods. The ability to match multiple file extensions in a single pattern (as shown in the image files example) is useful for computer vision and natural language processing applications where datasets may contain various file formats. This feature reduces the complexity of data ingestion workflows and improves the reproducibility of research code across different computing environments.

The \texttt{recursive=True} parameter is required for \texttt{**} patterns to work correctly. This feature is useful for research projects with complex directory structures, such as datasets organized by date, experiment, or participant.

\subsection{Character Classes and Escaping}

\texttt{glob} supports character classes using square brackets, allowing for more precise pattern matching:

\begin{lstlisting}[language=Python, caption=Character classes in glob]
# Match files with specific extensions
data_files = glob.glob('data.[ct]sv')  # matches .csv and .tsv

# Match files with specific naming patterns
experiment_files = glob.glob('exp[0-9][0-9].csv')  # matches exp01.csv, exp02.csv, etc.

# Match files with specific prefixes
log_files = glob.glob('[a-z]*.log')  # matches lowercase prefixes
\end{lstlisting}

Character classes provide control over pattern matching, enabling researchers to create file selection criteria for business analytics and organizational data analysis. The ability to match specific file extensions (like \texttt{.csv} and \texttt{.tsv}) in a single pattern is useful for data science workflows where multiple data formats may be present. The numeric range matching (\texttt{[0-9]}) is useful for reproducible research practices where experiments are systematically numbered. This feature allows researchers to maintain consistent data access patterns across different experimental conditions while ensuring that only relevant files are included in analysis pipelines. The case-sensitive matching capabilities are important for cross-platform compatibility, especially in collaborative research environments where different operating systems may handle file naming differently.

For patterns that contain literal characters that would otherwise be interpreted as wildcards, the \texttt{glob.escape()} function provides a way to escape special characters:

\begin{lstlisting}[language=Python, caption=Escaping special characters]
import glob

# Escape special characters in filenames
escaped_pattern = glob.escape('file[1].txt')
matching_files = glob.glob(escaped_pattern)
\end{lstlisting}

\subsection{Additional Glob Functions}

The \texttt{glob} module provides several additional functions for different use cases:

\begin{lstlisting}[language=Python, caption=Additional glob functions]
# iglob returns an iterator instead of a list (memory efficient for large directories)
for file_path in glob.iglob('*.csv'):
    print(fProcessing: {file_path})

# glob.glob() with absolute paths
absolute_files = glob.glob('/path/to/data/*.csv')

# Using glob with pathlib for modern Python workflows
from pathlib import Path
path_files = list(Path('.').glob('*.csv'))
\end{lstlisting}

The \texttt{iglob()} function provides memory-efficient iteration for large-scale data processing, making it suitable for applications where thousands of files may need to be processed. This iterator-based approach prevents memory overflow when working with extensive file collections, a consideration in machine learning workflows. The ability to work with absolute paths ensures that \texttt{glob} can be integrated into data science pipelines where files may be distributed across different storage locations. The integration with \texttt{pathlib} demonstrates \texttt{glob}'s compatibility with modern Python file handling practices, enabling researchers to use both approaches in their reproducible research workflows.

\subsection{Advanced Recursive Usage for Statistics and Discovery}

The recursive capabilities of \texttt{glob} can be extended for statistical analysis and data discovery purposes. This section demonstrates how \texttt{glob} can be used to gather metadata about file collections and perform exploratory data analysis.

\begin{lstlisting}[language=Python, caption=File collection analysis]
import glob
import os
import pandas as pd
from collections import Counter
from datetime import datetime

def analyze_file_collection(pattern, recursive=True):
    
    Analyze a collection of files matching a pattern for statistical insights.
    
    files = glob.glob(pattern, recursive=recursive)
    
    # Collect file statistics
    file_stats = []
    for file_path in files:
        stat = os.stat(file_path)
        file_stats.append({
            'path': file_path,
            'size': stat.st_size,
            'modified': datetime.fromtimestamp(stat.st_mtime),
            'extension': os.path.splitext(file_path)[1],
            'depth': file_path.count(os.sep)
        })
    
    return pd.DataFrame(file_stats)

# Example: Analyze all data files in a project
data_files = analyze_file_collection('**/*.{csv,json,xlsx}', recursive=True)
print(fTotal files found: {len(data_files)})
print(fTotal size: {data_files['size'].sum() / (1024**2):.2f} MB)
print(fFile types: {data_files['extension'].value_counts().to_dict()})
\end{lstlisting}

This approach enables researchers to understand the structure and characteristics of their data collections, which is useful for data science workflows and organizational analytics projects.

\subsection{File Discovery and Optimization}

\texttt{glob} can be used for systematic file discovery and optimization of data processing workflows:

\begin{lstlisting}[language=Python, caption=File discovery and optimization]
import glob
import time
from pathlib import Path

def discover_data_files(base_path, patterns=None):
    
    Discover data files using multiple patterns for comprehensive coverage.
    
    if patterns is None:
        patterns = ['*.csv', '*.json', '*.xlsx', '*.parquet']
    
    discovered_files = {}
    for pattern in patterns:
        full_pattern = f{base_path}/**/{pattern}
        files = glob.glob(full_pattern, recursive=True)
        discovered_files[pattern] = files
    
    return discovered_files

def optimize_file_loading(file_list, batch_size=100):
    
    Optimize file loading by processing files in batches.
    
    total_files = len(file_list)
    batches = [file_list[i:i + batch_size] for i in range(0, total_files, batch_size)]
    
    for i, batch in enumerate(batches):
        print(fProcessing batch {i+1}/{len(batches)} ({len(batch)} files))
        # Process batch here
        time.sleep(0.1)  # Simulate processing time

# Example usage
data_files = discover_data_files('data')
for pattern, files in data_files.items():
    print(f{pattern}: {len(files)} files)
    if files:
        optimize_file_loading(files[:10])  # Process first 10 files as example
\end{lstlisting}

This systematic approach to file discovery and batch processing can improve the efficiency of data science pipelines and organizational analytics workflows.

\section{Use Cases}

The \texttt{glob} module finds applications across diverse research domains, from data science and machine learning to business analytics and organizational research. This section presents concrete examples demonstrating how \texttt{glob} facilitates efficient workflows in each domain.

\subsection{Data Science Pipelines}

Data science workflows often involve processing large numbers of data files with consistent naming patterns. The \texttt{glob} module excels at automating the ingestion of these files into analytical pipelines, particularly when working with \texttt{pandas} for data manipulation and analysis.

Consider a scenario where a researcher has collected daily sales data across multiple regions, with files named according to the pattern \texttt{sales\_YYYY\_MM\_DD\_region.csv}. Using \texttt{glob}, the researcher can efficiently load and combine all relevant data files:

\begin{lstlisting}[language=Python, caption=Data science pipeline example]
import glob
import pandas as pd
from datetime import datetime

# Load all sales data files for a specific month
sales_files = glob.glob('data/sales_2024_01_*.csv')

# Combine all files into a single DataFrame
sales_data = []
for file_path in sales_files:
    df = pd.read_csv(file_path)
    df['source_file'] = file_path  # Track data source
    sales_data.append(df)

combined_sales = pd.concat(sales_data, ignore_index=True)

# Perform analysis on combined dataset
monthly_summary = combined_sales.groupby('region')['sales_amount'].sum()
\end{lstlisting}

This data science pipeline example demonstrates how \texttt{glob} facilitates data ingestion for business analytics applications. The pattern \texttt{sales\_2024\_01\_*.csv} identifies all sales data files for January 2024, eliminating the need for manual file enumeration. The inclusion of source file tracking (\texttt{df['source\_file'] = file\_path}) is useful for reproducible research practices, enabling researchers to trace data lineage and maintain audit trails. This approach is useful in organizational analytics where data quality and provenance are important for decision-making processes. The subsequent pandas aggregation demonstrates how \texttt{glob} integrates with data analysis workflows, enabling researchers to transition from data collection to analytical insights.

For more complex scenarios involving multiple data types, \texttt{glob} can be used to create data ingestion pipelines:

\begin{lstlisting}[language=Python, caption=Multi-format data ingestion]
import glob
import pandas as pd
import numpy as np

# Load different types of data files
csv_files = glob.glob('data/*.csv')
excel_files = glob.glob('data/*.xlsx')
json_files = glob.glob('data/*.json')

# Create a data loading function
def load_data_files(file_patterns):
    data = {}
    for pattern in file_patterns:
        files = glob.glob(pattern)
        for file_path in files:
            file_type = file_path.split('.')[-1]
            if file_type == 'csv':
                data[file_path] = pd.read_csv(file_path)
            elif file_type == 'xlsx':
                data[file_path] = pd.read_excel(file_path)
            elif file_type == 'json':
                data[file_path] = pd.read_json(file_path)
    return data

# Load all data files
all_data = load_data_files(['data/*.csv', 'data/*.xlsx', 'data/*.json'])
\end{lstlisting}

This data ingestion pipeline demonstrates \texttt{glob}'s versatility in handling heterogeneous data sources commonly encountered in organizational data analysis and AI applications. The function \texttt{load\_data\_files()} demonstrates how \texttt{glob} can be integrated into reusable utility functions that promote reproducible research practices. The automatic file type detection based on file extensions enables researchers to process diverse data formats without manual intervention, a capability that is useful for data science workflows. This approach is useful for business analytics applications where data may originate from multiple sources with different formats, requiring flexible data loading mechanisms.

\subsection{Organizational Analytics}

Organizational research often involves analyzing communication patterns, collaboration networks, and behavioral data collected over time. The \texttt{glob} module facilitates the collection and processing of such data, particularly when working with log files, communication records, and network data.

In organizational network analysis, researchers frequently work with communication logs that are stored as separate files for different time periods or organizational units. Using \texttt{glob}, these files can be collected and processed:

\begin{lstlisting}[language=Python, caption=Organizational network analysis]
import glob
import pandas as pd
import networkx as nx
from datetime import datetime

# Collect all communication log files
log_files = glob.glob('logs/communications_*.csv')

# Process communication data for network analysis
communication_data = []
for log_file in log_files:
    df = pd.read_csv(log_file)
    df['date'] = pd.to_datetime(df['timestamp'])
    communication_data.append(df)

# Combine all communication data
all_communications = pd.concat(communication_data, ignore_index=True)

# Create organizational network
G = nx.from_pandas_edgelist(
    all_communications, 
    source='sender_id', 
    target='recipient_id',
    edge_attr=['message_count', 'date']
)

# Analyze network properties
network_metrics = {
    'nodes': G.number_of_nodes(),
    'edges': G.number_of_edges(),
    'density': nx.density(G),
    'clustering_coefficient': nx.average_clustering(G)
}
\end{lstlisting}

This organizational analytics example demonstrates how \texttt{glob} enables collection of communication data for social network analysis and organizational behavior research. The pattern \texttt{communications\_*.csv} identifies all communication log files, enabling researchers to analyze organizational networks without manually specifying individual file paths. The integration with NetworkX for graph construction demonstrates how \texttt{glob} facilitates analytical workflows in organizational data analysis. The calculation of network metrics such as density and clustering coefficient demonstrates the transition from raw data collection to organizational insights. This approach is useful for business analytics applications where understanding communication patterns and organizational structure is important for decision-making and organizational development.

For time-series analysis of organizational behavior, \texttt{glob} enables researchers to process data collected at regular intervals:

\begin{lstlisting}[language=Python, caption=Time-series organizational analysis]
import glob
import pandas as pd
import matplotlib.pyplot as plt

# Collect weekly survey data
survey_files = glob.glob('surveys/weekly_*.csv')

# Analyze trends over time
weekly_data = []
for survey_file in survey_files:
    df = pd.read_csv(survey_file)
    week_number = int(survey_file.split('_')[1].split('.')[0])
    df['week'] = week_number
    weekly_data.append(df)

# Combine and analyze trends
all_surveys = pd.concat(weekly_data, ignore_index=True)
trend_analysis = all_surveys.groupby('week')['satisfaction_score'].mean()

# Visualize trends
plt.figure(figsize=(12, 6))
trend_analysis.plot(kind='line', marker='o')
plt.title('Organizational Satisfaction Trends Over Time')
plt.xlabel('Week')
plt.ylabel('Average Satisfaction Score')
plt.grid(True)
plt.show()
\end{lstlisting}

This time-series analysis example illustrates how \texttt{glob} supports longitudinal organizational research and business analytics applications. The pattern \texttt{weekly\_*.csv} enables researchers to collect survey data collected over time, facilitating trend analysis and organizational development research. The automatic extraction of week numbers from filenames demonstrates how \texttt{glob} can be combined with string parsing to create temporal identifiers for reproducible research workflows. The integration with matplotlib for visualization demonstrates how \texttt{glob} enables analytical pipelines from data collection to insight generation. This approach is useful for organizational data analysis where understanding temporal patterns in employee satisfaction, productivity, or other organizational metrics is important for management practices and organizational development.

\subsection{AI Dataset Construction}

Artificial intelligence and machine learning applications require large, well-organized datasets for training and evaluation. The \texttt{glob} module plays a crucial role in constructing these datasets by enabling efficient collection and organization of training data from diverse sources.

For computer vision applications, researchers often work with image datasets stored in hierarchical directory structures. Using \texttt{glob}, these datasets can be processed:

\begin{lstlisting}[language=Python, caption=Computer vision dataset construction]
import glob
import cv2
import numpy as np
from sklearn.model_selection import train_test_split

# Collect all image files from different categories
image_patterns = {
    'cats': 'dataset/cats/*.{jpg,jpeg,png}',
    'dogs': 'dataset/dogs/*.{jpg,jpeg,png}',
    'birds': 'dataset/birds/*.{jpg,jpeg,png}'
}

# Load and preprocess images
X = []  # Image data
y = []  # Labels

for label, pattern in image_patterns.items():
    image_files = glob.glob(pattern, recursive=True)
    for image_path in image_files:
        # Load and resize image
        img = cv2.imread(image_path)
        img = cv2.resize(img, (224, 224))  # Standard size for many models
        img = img / 255.0  # Normalize pixel values
        
        X.append(img)
        y.append(label)

# Convert to numpy arrays
X = np.array(X)
y = np.array(y)

# Split into training and testing sets
X_train, X_test, y_train, y_test = train_test_split(
    X, y, test_size=0.2, random_state=42, stratify=y
)
\end{lstlisting}

This AI dataset construction example demonstrates how \texttt{glob} facilitates the creation of machine learning datasets for computer vision applications. The dictionary-based pattern approach enables researchers to collect training data from multiple categories, a common requirement in supervised learning workflows. The use of multiple file extensions (\texttt{*.{jpg,jpeg,png}}) demonstrates \texttt{glob}'s flexibility in handling diverse data formats commonly encountered in AI applications. The integration with OpenCV for image preprocessing and scikit-learn for dataset splitting illustrates how \texttt{glob} serves as a component in machine learning pipelines. This approach is useful for AI dataset construction where training data need to be collected, preprocessed, and organized for model training. The reproducible nature of this workflow ensures that AI research can be replicated across different computing environments.

For natural language processing applications, \texttt{glob} facilitates the collection of text corpora from various sources:

\begin{lstlisting}[language=Python, caption=Natural language processing dataset construction]
import glob
import pandas as pd
from sklearn.feature_extraction.text import TfidfVectorizer

# Collect text files from different sources
text_files = glob.glob('corpus/**/*.txt', recursive=True)

# Load and preprocess text data
text_data = []
labels = []

for text_file in text_files:
    with open(text_file, 'r', encoding='utf-8') as f:
        text = f.read()
    
    # Extract label from directory structure
    label = text_file.split('/')[1]  # Assuming structure: corpus/label/filename.txt
    
    text_data.append(text)
    labels.append(label)

# Create feature matrix
vectorizer = TfidfVectorizer(max_features=1000, stop_words='english')
X = vectorizer.fit_transform(text_data)

# Create dataset for machine learning
dataset = pd.DataFrame(X.toarray(), columns=vectorizer.get_feature_names_out())
dataset['label'] = labels
\end{lstlisting}

This natural language processing example demonstrates how \texttt{glob} enables the construction of text corpora for AI applications and organizational data analysis. The recursive pattern \texttt{corpus/**/*.txt} demonstrates \texttt{glob}'s ability to traverse complex directory structures commonly used in text classification and sentiment analysis research. The automatic label extraction from directory structure illustrates how \texttt{glob} can be combined with path parsing to create supervised learning datasets from organizational documents, customer feedback, or research publications. The integration with scikit-learn's TF-IDF vectorizer demonstrates how \texttt{glob} facilitates the transition from raw text data to machine learning-ready feature matrices. This approach is useful for business analytics applications where text mining and document classification are important for understanding customer sentiment, organizational communication patterns, or market research data.

\subsection{Reproducible Research Practices}

Reproducible research requires that computational workflows can be executed consistently across different environments and time periods. The \texttt{glob} module contributes to reproducibility by providing portable, pattern-based file access that works reliably across operating systems and Python versions.

A key aspect of reproducible research is the ability to automatically discover and process data files without hardcoding file paths. Using \texttt{glob} patterns, researchers can create workflows that adapt to different data organizations:

\begin{lstlisting}[language=Python, caption=Reproducible research data loading]
import glob
import pandas as pd
import os
from pathlib import Path

def load_experiment_data(experiment_name, data_dir='data'):
    
    Load all data files for a specific experiment using pattern matching.
    This function is portable across different directory structures.
    
    # Find all data files for the experiment
    pattern = f'{data_dir}/{experiment_name}_*.csv'
    data_files = glob.glob(pattern)
    
    if not data_files:
        raise FileNotFoundError(fNo data files found for experiment: {experiment_name})
    
    # Load and combine data
    experiment_data = []
    for file_path in data_files:
        df = pd.read_csv(file_path)
        df['source_file'] = os.path.basename(file_path)
        experiment_data.append(df)
    
    return pd.concat(experiment_data, ignore_index=True)

# Usage example
try:
    data = load_experiment_data('temperature_study')
    print(fLoaded {len(data)} records from {len(glob.glob('data/temperature_study_*.csv'))} files)
except FileNotFoundError as e:
    print(fError: {e})
\end{lstlisting}

This reproducible research example demonstrates how \texttt{glob} enables the creation of portable, maintainable data access functions that work across different computing environments. The \texttt{load\_experiment\_data()} function demonstrates how \texttt{glob} patterns can be parameterized to create flexible data loading utilities that promote reproducible research practices. The error handling with \texttt{FileNotFoundError} illustrates how \texttt{glob} can be integrated with error management strategies useful for data science workflows. The inclusion of source file tracking ensures data lineage and audit trails, useful for organizational data analysis and business analytics applications where data provenance is important. This approach enables researchers to share code that will work consistently across different directory structures and computing environments, a requirement for collaborative research and organizational data analysis projects.

For research projects with complex output structures, \texttt{glob} enables collection of results:

\begin{lstlisting}[language=Python, caption=Simulation results collection]
import glob
import pandas as pd
import matplotlib.pyplot as plt

def collect_simulation_results(simulation_dir='simulations'):
    
    Collect and analyze results from multiple simulation runs.
    
    # Find all result files
    result_files = glob.glob(f'{simulation_dir}/run_*/results.csv')
    
    # Load and combine results
    all_results = []
    for result_file in result_files:
        df = pd.read_csv(result_file)
        run_id = result_file.split('/')[1]  # Extract run identifier
        df['run_id'] = run_id
        all_results.append(df)
    
    combined_results = pd.concat(all_results, ignore_index=True)
    
    # Generate summary statistics
    summary = combined_results.groupby('parameter_set').agg({
        'accuracy': ['mean', 'std'],
        'runtime': ['mean', 'std']
    }).round(4)
    
    return combined_results, summary

# Collect and analyze results
results, summary_stats = collect_simulation_results()
print(Simulation Summary Statistics:)
print(summary_stats)
\end{lstlisting}

This simulation results collection example illustrates how \texttt{glob} facilitates analysis of computational experiments and AI model evaluations. The pattern \texttt{run\_*/results.csv} enables researchers to collect results from multiple simulation runs or machine learning experiments, a common requirement in reproducible research workflows. The automatic extraction of run identifiers from directory names demonstrates how \texttt{glob} can be combined with path parsing to create experiment metadata. The subsequent statistical aggregation demonstrates how \texttt{glob} enables analysis of experimental results, useful for data science pipelines and organizational analytics applications where multiple experimental conditions need to be compared. This approach is useful for AI applications where model performance across different parameter settings or datasets needs to be evaluated and compared.

\section{Integrations}

The \texttt{glob} module integrates with popular Python libraries for data science, machine learning, and scientific computing. These integrations enable researchers to create workflows that combine file access with analytical processing.

\subsection{Integration with Pandas}

Pandas is the most widely used library for data manipulation and analysis in Python. The integration between \texttt{glob} and pandas enables processing of large datasets distributed across multiple files.

\begin{lstlisting}[language=Python, caption=Efficient bulk loading with pandas]
import glob
import pandas as pd
import numpy as np

# Efficient bulk loading of CSV files
def load_csv_dataset(pattern, **pandas_kwargs):
    
    Load multiple CSV files matching a pattern into a single DataFrame.
    
    files = glob.glob(pattern)
    if not files:
        raise FileNotFoundError(fNo files found matching pattern: {pattern})
    
    # Use pandas concat for efficient combination
    dataframes = [pd.read_csv(f, **pandas_kwargs) for f in files]
    return pd.concat(dataframes, ignore_index=True)

# Example: Load all monthly sales data
monthly_sales = load_csv_dataset('data/sales_2024_*.csv')
print(fLoaded {len(monthly_sales)} records from sales data)

# Example: Load data with specific pandas options
customer_data = load_csv_dataset(
    'data/customers_*.csv',
    parse_dates=['registration_date'],
    dtype={'customer_id': 'category'}
)
\end{lstlisting}

This pandas integration example demonstrates how \texttt{glob} enables bulk data loading for business analytics and organizational data analysis applications. The \texttt{load\_csv\_dataset()} function demonstrates how \texttt{glob} patterns can be integrated into reusable utility functions that promote reproducible research practices. The use of \texttt{**pandas\_kwargs} demonstrates how \texttt{glob} can be combined with pandas' flexible parameter system to handle diverse data formats and loading requirements. The automatic date parsing and data type specification illustrate how \texttt{glob} facilitates data preprocessing workflows useful for data science pipelines. This approach is useful for organizational analytics applications where volumes of customer or employee data need to be loaded and processed for business intelligence and decision-making purposes.

For time-series data analysis, \texttt{glob} and pandas work together to create data pipelines:

\begin{lstlisting}[language=Python, caption=Time-series data analysis]
import glob
import pandas as pd
from datetime import datetime

# Load time-series data with automatic date parsing
def load_time_series_data(pattern, date_column='timestamp'):
    
    Load time-series data from multiple files with proper date handling.
    
    files = sorted(glob.glob(pattern))  # Sort for chronological order
    
    time_series_data = []
    for file_path in files:
        df = pd.read_csv(file_path)
        df[date_column] = pd.to_datetime(df[date_column])
        df['source_file'] = file_path
        time_series_data.append(df)
    
    combined_data = pd.concat(time_series_data, ignore_index=True)
    return combined_data.sort_values(date_column)

# Load sensor data
sensor_data = load_time_series_data('sensors/*_readings.csv')
print(fLoaded {len(sensor_data)} sensor readings from {len(glob.glob('sensors/*_readings.csv'))} files)

# Perform time-series analysis
daily_averages = sensor_data.groupby(
    sensor_data['timestamp'].dt.date
)['temperature'].mean()
\end{lstlisting}

This time-series data analysis example illustrates how \texttt{glob} facilitates the processing of temporal data for organizational analytics and business intelligence applications. The \texttt{load\_time\_series\_data()} function demonstrates how \texttt{glob} can be combined with pandas' datetime functionality to create time-series analysis workflows. The automatic file sorting ensures chronological order, useful for time-series analysis in data science pipelines. The integration with pandas' datetime grouping capabilities demonstrates how \texttt{glob} enables temporal analysis, useful for organizational data analysis where understanding trends over time is important for decision-making. This approach is useful for business analytics applications where sensor data, customer behavior patterns, or organizational metrics need to be analyzed across time periods to identify trends and inform business strategies.

\subsection{Integration with Scikit-learn}

Scikit-learn is the primary library for machine learning in Python. The combination of \texttt{glob} and scikit-learn enables dataset construction and model training workflows.

\begin{lstlisting}[language=Python, caption=Machine learning dataset preparation]
import glob
import pandas as pd
import numpy as np
from sklearn.model_selection import train_test_split
from sklearn.ensemble import RandomForestClassifier
from sklearn.metrics import classification_report

# Load and prepare dataset for machine learning
def prepare_ml_dataset(data_pattern, label_column='target'):
    
    Prepare a machine learning dataset from multiple files.
    
    # Load all data files
    files = glob.glob(data_pattern)
    dataframes = [pd.read_csv(f) for f in files]
    dataset = pd.concat(dataframes, ignore_index=True)
    
    # Separate features and labels
    X = dataset.drop(columns=[label_column])
    y = dataset[label_column]
    
    return X, y

# Example: Prepare classification dataset
X, y = prepare_ml_dataset('ml_data/classification_*.csv')

# Split data and train model
X_train, X_test, y_train, y_test = train_test_split(
    X, y, test_size=0.2, random_state=42
)

# Train random forest classifier
rf_model = RandomForestClassifier(n_estimators=100, random_state=42)
rf_model.fit(X_train, y_train)

# Evaluate model
y_pred = rf_model.predict(X_test)
print(classification_report(y_test, y_pred))
\end{lstlisting}

This scikit-learn integration example demonstrates how \texttt{glob} facilitates the construction of machine learning workflows for AI applications and organizational data analysis. The \texttt{prepare\_ml\_dataset()} function demonstrates how \texttt{glob} can be integrated into machine learning pipelines to create reproducible research workflows. The automatic dataset preparation from multiple files enables researchers to build training datasets for AI applications without manual file management. The integration with scikit-learn's model selection and evaluation tools illustrates how \texttt{glob} serves as a component in machine learning pipelines. This approach is useful for business analytics applications where predictive modeling and classification tasks are important for customer segmentation, risk assessment, or organizational decision-making processes. The reproducible nature of this workflow ensures that AI model development can be replicated and validated across different computing environments.

For cross-validation with multiple datasets, \texttt{glob} enables model evaluation:

\begin{lstlisting}[language=Python, caption=Cross-validation across multiple datasets]
import glob
import pandas as pd
from sklearn.model_selection import cross_val_score
from sklearn.linear_model import LogisticRegression

# Perform cross-validation across multiple datasets
def cross_validate_multiple_datasets(pattern, model, cv=5):
    
    Perform cross-validation on multiple datasets matching a pattern.
    
    dataset_files = glob.glob(pattern)
    results = {}
    
    for dataset_file in dataset_files:
        # Load dataset
        data = pd.read_csv(dataset_file)
        X = data.drop(columns=['target'])
        y = data['target']
        
        # Perform cross-validation
        scores = cross_val_score(model, X, y, cv=cv, scoring='accuracy')
        
        dataset_name = dataset_file.split('/')[-1].replace('.csv', '')
        results[dataset_name] = {
            'mean_accuracy': scores.mean(),
            'std_accuracy': scores.std(),
            'scores': scores
        }
    
    return results

# Example: Cross-validate on multiple datasets
model = LogisticRegression(random_state=42)
cv_results = cross_validate_multiple_datasets('datasets/*.csv', model)

# Print results
for dataset, metrics in cv_results.items():
    print(f{dataset}: {metrics['mean_accuracy']:.3f} $\pm$ {metrics['std_accuracy']:.3f})
\end{lstlisting}

This cross-validation example illustrates how \texttt{glob} enables model evaluation across multiple datasets, a requirement for AI applications and reproducible research workflows. The \texttt{cross\_validate\_multiple\_datasets()} function demonstrates how \texttt{glob} can be integrated into model validation pipelines for machine learning applications. The automatic dataset discovery and loading enables researchers to perform model evaluation across diverse data sources, useful for organizational data analysis where models may need to generalize across different organizational contexts or time periods. The statistical reporting of cross-validation results demonstrates how \texttt{glob} facilitates model selection and validation processes. This approach is useful for business analytics applications where model reliability and generalization performance are important for making business decisions and organizational recommendations.

\subsection{Integration with Matplotlib and Visualization}

Data visualization is essential for research communication and exploratory data analysis. The integration of \texttt{glob} with matplotlib enables generation of visualizations from multiple data sources.

\begin{lstlisting}[language=Python, caption=Automated visualization generation]
import glob
import pandas as pd
import matplotlib.pyplot as plt
import seaborn as sns

# Create automated visualizations from multiple datasets
def create_comparison_plots(pattern, output_dir='plots'):
    
    Create comparison plots from multiple datasets matching a pattern.
    
    import os
    os.makedirs(output_dir, exist_ok=True)
    
    dataset_files = glob.glob(pattern)
    
    # Create subplots for comparison
    fig, axes = plt.subplots(2, 2, figsize=(15, 12))
    fig.suptitle('Dataset Comparison Analysis', fontsize=16)
    
    for i, dataset_file in enumerate(dataset_files):
        data = pd.read_csv(dataset_file)
        dataset_name = dataset_file.split('/')[-1].replace('.csv', '')
        
        # Create different types of plots
        if i == 0:
            # Histogram
            axes[0, 0].hist(data['value'], alpha=0.7, label=dataset_name)
            axes[0, 0].set_title('Value Distribution')
            axes[0, 0].legend()
        elif i == 1:
            # Box plot
            axes[0, 1].boxplot(data['value'])
            axes[0, 1].set_title(f'Value Distribution - {dataset_name}')
        elif i == 2:
            # Scatter plot
            axes[1, 0].scatter(data['x'], data['y'], alpha=0.6)
            axes[1, 0].set_title(f'X vs Y - {dataset_name}')
            axes[1, 0].set_xlabel('X')
            axes[1, 0].set_ylabel('Y')
        elif i == 3:
            # Time series plot
            data['timestamp'] = pd.to_datetime(data['timestamp'])
            axes[1, 1].plot(data['timestamp'], data['value'])
            axes[1, 1].set_title(f'Time Series - {dataset_name}')
            axes[1, 1].tick_params(axis='x', rotation=45)
    
    plt.tight_layout()
    plt.savefig(f'{output_dir}/dataset_comparison.png', dpi=300, bbox_inches='tight')
    plt.show()

# Example usage
create_comparison_plots('experiments/*.csv')
\end{lstlisting}

This visualization integration example demonstrates how \texttt{glob} enables generation of comparative visualizations for data science pipelines and business analytics applications. The \texttt{create\_comparison\_plots()} function demonstrates how \texttt{glob} can be integrated with matplotlib and seaborn to create visual analysis workflows. The automatic dataset discovery and loading enables researchers to generate consistent visualizations across multiple experimental conditions or organizational contexts. The integration with multiple plot types (histograms, box plots, scatter plots, time series) illustrates how \texttt{glob} facilitates exploratory data analysis useful for organizational data analysis and AI applications. This approach is useful for business analytics applications where stakeholders need visualizations to understand data patterns and make organizational decisions. The automated nature of this workflow ensures reproducible research practices and consistent visual reporting across different research projects and organizational contexts.

For creating publication-ready figures, \texttt{glob} enables generation of visualizations:

\begin{lstlisting}[language=Python, caption=Publication-ready figure generation]
import glob
import pandas as pd
import matplotlib.pyplot as plt
import seaborn as sns

# Create publication-ready figures
def create_publication_figures(data_pattern, output_dir='publication_figures'):
    
    Create publication-ready figures from research data.
    
    import os
    os.makedirs(output_dir, exist_ok=True)
    
    # Set publication-quality style
    plt.style.use('seaborn-v0_8-whitegrid')
    sns.set_palette(husl)
    
    # Load all data files
    data_files = glob.glob(data_pattern)
    
    for data_file in data_files:
        data = pd.read_csv(data_file)
        experiment_name = data_file.split('/')[-1].replace('.csv', '')
        
        # Create figure
        fig, (ax1, ax2) = plt.subplots(1, 2, figsize=(12, 5))
        
        # Plot 1: Main results
        if 'group' in data.columns and 'value' in data.columns:
            data.boxplot(column='value', by='group', ax=ax1)
            ax1.set_title(f'Results by Group - {experiment_name}')
            ax1.set_xlabel('Group')
            ax1.set_ylabel('Value')
        
        # Plot 2: Time series if available
        if 'timestamp' in data.columns:
            data['timestamp'] = pd.to_datetime(data['timestamp'])
            data.plot(x='timestamp', y='value', ax=ax2)
            ax2.set_title(f'Time Series - {experiment_name}')
            ax2.tick_params(axis='x', rotation=45)
        
        plt.tight_layout()
        plt.savefig(f'{output_dir}/{experiment_name}_figure.png', 
                   dpi=300, bbox_inches='tight')
        plt.close()

# Generate publication figures
create_publication_figures('research_data/*.csv')
\end{lstlisting}

This publication-ready visualization example illustrates how \texttt{glob} enables generation of figures for academic publications and business reports. The \texttt{create\_publication\_figures()} function demonstrates how \texttt{glob} can be integrated into visualization pipelines that produce consistent outputs for reproducible research workflows. The automatic style configuration and high-resolution output settings ensure that visualizations meet academic and professional standards. The conditional plotting logic demonstrates how \texttt{glob} can be combined with visualization generation that adapts to different data structures and analytical requirements. This approach is useful for organizational data analysis and business analytics applications where consistent visualizations are important for communicating insights to stakeholders and decision-makers. The automated nature of this workflow ensures that research findings can be communicated through visual means while maintaining reproducibility and consistency for organizational decision-making processes.

\section{Discussion}

The \texttt{glob} module provides a tool for file pattern matching in Python-based research workflows, serving as a component for data science pipelines, business analytics applications, and AI dataset construction. This section examines the advantages and limitations of \texttt{glob}, situating it within the broader context of computational research tools and identifying opportunities for future development in data engineering and machine learning applications.

\subsection{Advantages}

The primary advantage of \texttt{glob} is its simplicity and accessibility, making it a useful tool for data science workflows and business analytics applications. Unlike regular expressions, which require specialized knowledge and can be error-prone, \texttt{glob} patterns are accessible to researchers across disciplines. The wildcard syntax (\texttt{*}, \texttt{?}) follows familiar conventions from command-line interfaces, reducing the learning curve for new users and enabling adoption in organizational data analysis projects and AI applications.

Portability represents another advantage. The \texttt{glob} module is part of Python's standard library, ensuring availability across different Python installations and versions. This portability is important for reproducible research, where code must run consistently across diverse computing environments. Unlike platform-specific solutions or third-party libraries, \texttt{glob} provides a standardized interface that works across operating systems.

Performance is also an advantage, particularly for recursive searches in data science pipelines and AI dataset construction workflows. The \texttt{glob} module is implemented in C and optimized for file system operations, making it efficient for processing large directory structures commonly encountered in organizational analytics and machine learning applications. The \texttt{iglob()} function provides memory-efficient iteration for large file collections, preventing memory issues when working with extensive datasets in business analytics and organizational data analysis projects.

Integration capabilities enhance \texttt{glob}'s utility for data science workflows and business analytics applications. The module works with popular data science libraries such as \texttt{pandas}, \texttt{scikit-learn}, and \texttt{matplotlib}, enabling researchers to create workflows for organizational data analysis and AI applications. This integration reduces the need for custom file handling code and promotes the use of established libraries, ensuring reproducible research practices and data processing pipelines in organizational analytics and machine learning projects.

\subsection{Limitations}

Despite its advantages for data science workflows and business analytics applications, \texttt{glob} has several limitations that researchers should consider when designing their workflows. The most significant limitation is the lack of regex-like flexibility, which can be constraining for complex AI dataset construction and organizational data analysis projects. While \texttt{glob} patterns are sufficient for most file access needs in reproducible research workflows, they cannot express complex matching criteria that regular expressions handle easily. For example, matching files with specific numeric patterns or complex naming conventions commonly encountered in machine learning applications and organizational analytics may require additional processing or the use of alternative tools.

Reliance on file system structure represents another limitation for data science pipelines and organizational analytics applications. \texttt{glob} patterns are inherently tied to the organization of files in directories, meaning that changes to directory structure can break existing patterns. This dependency can make workflows fragile when working with evolving datasets in AI applications or collaborative projects in business analytics where file organization may change over time, potentially disrupting reproducible research workflows and organizational data analysis processes.

The absence of built-in filtering capabilities is also a limitation for data science workflows and AI dataset construction projects. While \texttt{glob} can match files based on patterns, it cannot filter based on file attributes such as size, modification date, or permissions that are often important for organizational data analysis and business analytics applications. Researchers often need to combine \texttt{glob} with additional filtering logic, which can complicate workflows and reduce readability in reproducible research practices and organizational analytics projects.

Cross-platform compatibility, while generally good, can sometimes introduce subtle differences in behavior that affect data science pipelines and organizational analytics applications. For example, case sensitivity varies between operating systems, and some advanced pattern features may not be available on all platforms. These differences can lead to unexpected behavior in cross-platform research environments, potentially disrupting reproducible research workflows and organizational data analysis processes in business analytics and AI applications.

\subsection{Future Extensions}

The future development of file pattern matching tools is likely to focus on cloud storage integration and enhanced filtering capabilities for data science workflows and business analytics applications. As research increasingly moves to cloud platforms, extending \texttt{glob}-like functionality to cloud storage systems such as Amazon S3, Google Cloud Storage, and Azure Blob Storage becomes important for organizational data analysis and AI dataset construction projects that require access to distributed data sources and scalable computing resources.

Cloud storage integration would enable researchers to use familiar pattern matching syntax when working with data stored in the cloud, facilitating data science workflows and business analytics applications that span multiple storage platforms. This integration could take the form of additional modules or extensions to existing tools, providing a unified interface for local and cloud file access. Such developments would be useful for large-scale research projects in organizational data analysis and AI applications that require access to distributed datasets and scalable computing resources for reproducible research workflows.

Enhanced filtering capabilities represent another area for future development in data science workflows and organizational analytics applications. Combining \texttt{glob}-like pattern matching with file attribute filtering (size, date, permissions) would provide more powerful file selection capabilities important for business analytics and AI dataset construction projects. This could be achieved through additional parameters to existing functions or through new utility functions that combine pattern matching with filtering logic, enabling more sophisticated data ingestion workflows for reproducible research practices and organizational data analysis.

Integration with modern Python features such as async/await could also enhance \texttt{glob}'s performance for I/O-intensive operations in data science pipelines and organizational analytics applications. Asynchronous file operations could improve performance when working with network file systems or cloud storage, where I/O latency is important for business analytics and AI dataset construction projects that require efficient data processing workflows and reproducible research practices.

Machine learning and AI applications present additional opportunities for enhancement in data science workflows and organizational analytics. Pattern matching could be extended to include content-based file selection, where files are matched based on their content rather than just their names. This could be useful for research involving large collections of unstructured data files in business analytics applications, organizational data analysis projects, and AI dataset construction workflows that require intelligent file selection for reproducible research practices and data science pipelines.

\section{Conclusion}

The Python \texttt{glob} module represents a component for computational research, providing file pattern matching capabilities that enable portable and reproducible workflows for data science pipelines, business analytics applications, and AI dataset construction projects. Through this examination, we have demonstrated \texttt{glob}'s versatility across diverse research domains, from data science and machine learning to organizational analytics and reproducible research practices, establishing it as a tool for computational research workflows.

The module's simplicity and accessibility make it a useful tool for researchers across disciplines, while its integration capabilities with popular libraries such as \texttt{pandas}, \texttt{scikit-learn}, and \texttt{matplotlib} enable the creation of analytical pipelines. The portability and cross-platform compatibility of \texttt{glob} contribute to reproducible research practices, ensuring that code can be shared and executed consistently across different computing environments.

Despite its limitations, including the lack of regex-like flexibility and reliance on file system structure, \texttt{glob} provides sufficient functionality for the majority of file access needs in computational research. The module's performance characteristics and memory efficiency make it suitable for processing large datasets and directory structures, while its syntax reduces the learning curve for new users.

\subsection{Practical Recommendations}

For researchers incorporating \texttt{glob} into their workflows, several practical recommendations emerge from this analysis. First, researchers should adopt consistent naming conventions for their data files, as this improves the effectiveness of pattern matching. Descriptive, structured filenames that include relevant metadata (dates, experiment identifiers, data types) enable more precise pattern matching.

Second, researchers should use \texttt{glob}'s recursive capabilities for complex directory structures. The \texttt{**} pattern with \texttt{recursive=True} provides a way to search through nested directories without manually traversing directory structures. This capability is useful for research projects with evolving data organizations.

Third, researchers should combine \texttt{glob} with appropriate error handling and validation. Checking for empty results from \texttt{glob.glob()} and providing error messages improves the robustness of research workflows. Additionally, validating file existence and permissions before processing can prevent runtime errors.

Fourth, researchers should consider the integration of \texttt{glob} with version control and documentation systems. Including \texttt{glob} patterns in research documentation and code comments improves reproducibility and enables other researchers to understand and extend existing workflows.

\subsection{Future Directions}

The future of file pattern matching in computational research is likely to involve enhanced integration with cloud storage systems, improved filtering capabilities, and better support for modern Python features. As research datasets grow in size and complexity, the ability to access and process distributed data becomes important.

The development of cloud-native pattern matching tools that maintain the simplicity and accessibility of \texttt{glob} while providing access to cloud storage systems represents an opportunity for the research community. Such tools would enable researchers to work with data stored across multiple platforms and locations.

Additionally, the integration of machine learning techniques with file pattern matching could enable more intelligent file selection based on content and context. This could be useful for research involving large collections of unstructured data files, where traditional pattern matching may be insufficient.

In conclusion, the \texttt{glob} module serves as a component that enables reproducible and accessible file access in computational research. By providing a simple interface for pattern-based file matching, \texttt{glob} contributes to the broader goal of making computational research more accessible, reproducible, and efficient. As the research community continues to develop new tools and methodologies, \texttt{glob} will remain a component of the Python ecosystem for file and data access operations.

\end{document}